\documentclass[pdflatex,sn-mathphys-num]{sn-jnl}

\usepackage{xr}

\usepackage{algorithm}%
\usepackage{algorithmicx}%
\usepackage{algpseudocode}%
\usepackage{amsmath,amssymb,amsfonts}%
\usepackage{amsthm}%
\usepackage[utf8]{inputenc}
\usepackage{anyfontsize}
\usepackage{float}
\usepackage[T1]{fontenc}
\usepackage[title]{appendix}%
\usepackage{booktabs}%
\usepackage{graphicx}%
\usepackage{listings}%
\usepackage{lmodern}
\usepackage{gensymb}
\usepackage{manyfoot}%
\usepackage{mathrsfs}%
\usepackage[version=4]{mhchem}
\usepackage{multirow}%
\usepackage{textcomp}
\usepackage{ulem}
\usepackage{units}
\usepackage{amssymb} 
\usepackage{xcolor}%
\hypersetup{linkcolor=black,citecolor=black,filecolor=black,urlcolor=black}

\usepackage{xr}
\theoremstyle{thmstyleone}%
\theoremstyle{thmstyletwo}%

\theoremstyle{thmstylethree}%

\begin{document}

\title[Article Title]{Precise one-dimensional nanochannels in transition metal dichalcogenides as building blocks for advanced nanophotonics}


\author*[1]{\fnm{Abhay V.} \sur{Agrawal}}\email{abhayv@chalmers.se}  
\author[2]{\fnm{Wouter} \sur{Holman}}
\author[1]{\fnm{Bet$\textrm{{\"ul}}$} \sur{K$\textrm{{\"u}}$$\textrm{{\c{c}}}$$\textrm{{\"u}}$k$\textrm{{\"o}}$z}}

\author[1,3]{\fnm{Tomasz J.} \sur{Antosiewicz}}

\author*[1]{\fnm{Timur O.} \sur{Shegai}}\email{timurs@chalmers.se}


\affil[1]{\orgdiv{Department of Physics and Astronomy}, \orgname{Chalmers University of Technology}, \postcode{412 96}, \city{G$\textrm{{\"o}}$teborg}, \country{Sweden}}

\affil[2]{\orgdiv{Department of Applied Physics and Science Education, Institute for Complex Molecular Systems, and
Eindhoven Hendrik Casimir Institute}, \orgname{ Eindhoven University of Technology}, \postcode{5600 MB}, \city{Eindhoven}, \country{Netherlands}}

\affil[3]{\orgdiv{Faculty of Physics}, \orgname{University of Warsaw}, \orgaddress{\street{Pasteura 5}, \postcode{02-093}, \city{Warsaw}, \country{Poland}}}

\abstract{Atomically sharp edges are essential for future high-index nanophotonic structures, yet conventional lithography and dry etching methods inevitably introduce edge roughness that limits optical confinement and reproducibility. Recently, anisotropic wet etching of multilayer van der Waals crystals, such as transition metal dichalcogenides (TMDs), has enabled crystallographically defined, atomically sharp zigzag edges, thereby eliminating the edge-roughness problem. However, the process is intrinsically limited to confined geometries such as isolated triangular or hexagonal features dictated by crystal stacking symmetry. Here, we demonstrate a lithography-guided anisotropic etching framework that drives TMDs etching beyond isolated confined geometries by enforcing controlled interaction of neighboring etched nanoholes regions. In multilayer 2H-\ce{WS2}, merging of anisotropic etch fronts enables sustained long-range propagation of zigzag facets, introducing a previously inaccessible 180$\degree$ edge alignment and a crystallographically defined design space combining 120$\degree$ and 180$\degree$ junctions. Using this approach, we fabricate proof-of-principle extended nanophotonic structures with ultrasharp sidewalls (approaching atomic precision), including sub-100-nm-gap one-dimensional gratings, waveguides, defect-engineered photonic cavities, angle-programmed photonic lattices, and radially patterned diffractive zone plates. Beyond structural characterization, back-focal-plane reflection spectroscopy of atomically sharp 1D periodic 2H-\ce{WS2} gratings demonstrates their photonic functionality, revealing symmetry-protected bound states in the continuum (SP-BICs) and strong exciton--photon coupling in multilayer \ce{WS2}. Finally, we fabricate ultrathin, ultranarrow, and ultralong nanoribbons with record-high aspect ratios. Together, these results demonstrate, to our knowledge for the first time, edge merging as a generic route to fabricate edge-defined, atomically sharp nanophotonic and nanoelectronic architectures in layered van der Waals platforms.}




\maketitle
\section{Introduction}

Fabrication of advanced nanophotonic devices has drawn much growing interest due to their exceptional use in nonlinear optics, optical trapping, sensing, communications, and in strong light-matter interactions~\cite{zotev2025nanophotonics,couteau2023applications,wang2016giant}. Most existing nanophotonic structures are manufactured using electron-beam lithography (EBL) followed by dry etching methods, which, while offering high resolution, often produce nanostructures with significant rough sidewall and edge roughness, structural imperfections due to proximity correction effects, fabrication tolerances, and resist pattern collapse issues~\cite{choi2025unidirectional,ban2023study,trugler2011influence}. These imperfections are especially critical in high-index structures such as gratings and waveguides, where even nanometer-scale deviations cause increased scattering, reduced diffraction efficiency, and lower $Q$-factors~\cite{yakuhina2020investigation,ban2023study,wang2022study,marciniak2021impact}. Such fabrication-induced roughness not only limits the optical efficiency but also restricts the scalability of nanophotonic integration. This emphasizes the urgent need for advanced methods capable of producing defect-free, smooth, and atomically sharp high-index nanostructures optimized for enhanced light–matter interactions. Integrating these photonic structures onto a chip with existing platforms such as silicon (Si), indium phosphide (InP), silicon nitride (\ce{Si3N4}), lithium niobate (\ce{LiNbO3}), noble metal-based plasmonic platforms, etc., remains challenging~\cite {leuthold2010nonlinear,poberaj2012lithium,hoefler2019foundry,moss2013new}. Silicon-based systems are constrained by their indirect bandgap, while hybrid integration is hampered by lattice mismatch. InP and \ce{LiNbO3} suffer from complementary metal-oxide-semiconductor (CMOS) incompatibility or low index contrast. Moreover, plasmonics-based systems have long faced performance bottlenecks arising from inherent absorption losses (and related to that excessive Joule heating), scarcity of high-quality plasmonic materials, and constraints on achievable nanoscale fabrication geometries~\cite{west2010searching,khurgin2015deal,baranov2017all}. These limitations have accelerated interest in alternative high-index dielectric materials such as transition metal dichalcogenides (TMDs), which, among other things, offer the possibility of atomically sharp nanophotonics~\cite{munkhbat2020transition}. 

Multilayer TMDs represent a unique class of high-index dielectric materials, featuring exceptional in-plane refractive index ($n>4$), low absorption in the near infrared region (NIR), and pronounced optical anisotropy~\cite{verre2019transition,zograf2024combining}. These characteristics make TMDs highly attractive for the development of future nanophotonic components such as gratings, waveguides, photonic cavities, and diffractive zone plates~\cite{munkhbat2020transition}. Because of weak van der Waals (vdW) interlayer bonding between layers, TMDs provide exceptional control over thickness down to the monolayer limit, enabling fabrication of ultra-smooth, lithographically defined nanophotonic structures~\cite{mak2010atomically}. In addition, one-dimensional (1D) edges in TMDs offer properties fundamentally different from those of the basal plane, which arises from reduced coordination, broken symmetry, and localized electronic states~\cite{pan2012edge,rossi2017effect}. These edges can have metallic or ferromagnetic states as demonstrated by density functional theory (DFT) and scanning tunneling microscopy (STM) studies~\cite{bollinger2001one,pan2012edge,xu2016oscillating}. Such edge-localized electronic bands support efficient charge transport and strong excitonic coupling, making them attractive for electrical, catalytic, and gas sensing applications~\cite{agrawal2017fast,polyakov2024top}. Moreover, the 1D edges of TMDs inherently support broken symmetry that leads to enhanced optical nonlinearities~\cite{zograf2025ultrathin,dewambrechies2023enhanced}. These localized states significantly amplify the optical response, producing sharp, edge-dominated enhanced nonlinear optical signals that outperform the basal plane contributions~\cite{yin2014edge,lin2018atom}. These insights highlight the necessity of engineering TMDs with well-defined edge features to accurately control light–matter interactions.

Edge-engineering in TMDs can be broadly approached through either top-down or bottom-up methods, each offering certain advantages and disadvantages~\cite{agrawal2018controlled,munkhbat2020transition,isoniemi2024realization}. In bottom-up approaches, chemical vapor deposition (CVD) remains the most widely used technique for growing wafer-scale TMDs; however, achieving deterministic control over the precise zigzag or armchair edge terminations is to date extremely challenging. Beyond the difficulty of controlling zigzag or armchair terminations, CVD-grown TMDs also suffer the complexity of achieving single-crystalline TMDs, uncontrolled grain boundaries, substrate variation, and incomplete chalcogenization, which altogether hinder the realization of high-quality materials with atomically defined edges for nanophotonics~\cite{chandler1993chemical,hossain2024recent}. These challenges can be effectively circumvented using top-down approaches based on mechanically exfoliated TMDs. The exfoliated flakes, derived directly from bulk single crystals, although suffering from scalability challenges, provide a preserved crystalline nature, negligible grain boundaries, and chemical purity, making them suitable for the fabrication of advanced nanophotonic structures on diverse substrate platforms.

To date, most TMD nanophotonic structures rely on top-down fabrication; however, the methods used, namely, EBL and subsequent dry etching, unavoidably introduce edge roughness, disorder, and sidewall damage that hinder atomically sharp boundaries~\cite{peng20253r,choi2025unidirectional,froch2019photonic,ban2023study,shen2022transition,pruszynska2025optical}. 

\begin{figure}
    \centering
    \includegraphics[width = 0.95\linewidth]{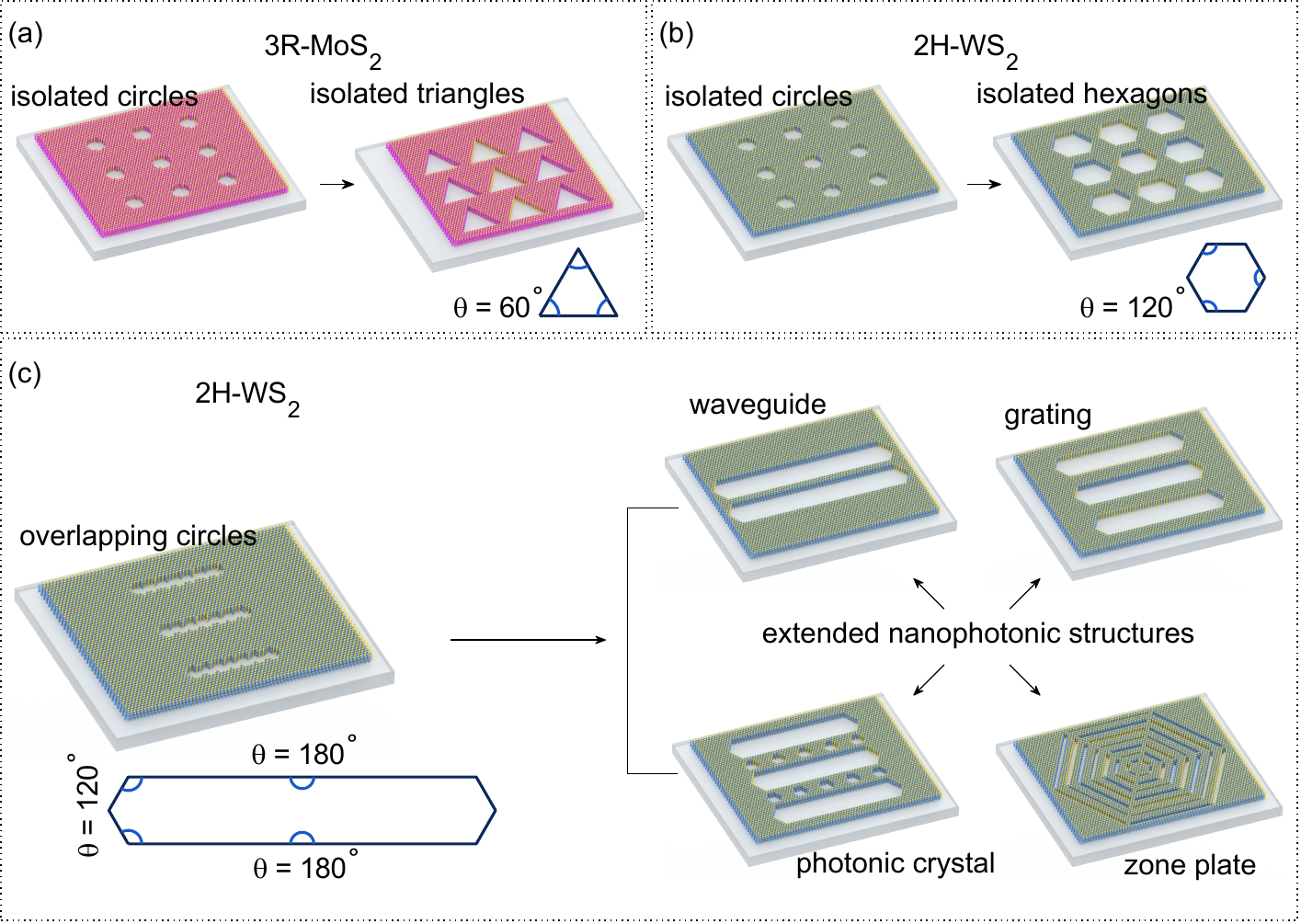}
    \caption{\textbf{Conceptual illustration of stacking-symmetry-dependent anisotropic etching and emergence of extended nanophotonic structures}: \textbf{a} Anisotropic wet etching of lithographically defined isolated nanoholes yields atomically sharp triangular features with 60\textdegree\ internal angles in multilayer 3R-TMDs. \textbf{b} Anisotropic wet etching of lithographically defined isolated nanoholes yields alternating zigzag-terminated hexagonal shapes with 120\textdegree\ internal angles in multilayer 2H-stacked TMDs. \textbf{c} Lithographically enforced alignment and spacing of nanoholes in multilayer 2H-TMDs enable interaction and merging of anisotropic etch fronts into continuous 180\textdegree\ aligned edges, allowing the formation of extended atomically sharp nanophotonic structures, including waveguides, gratings, and diffractive structures.}
    \label{fig:1}
\end{figure}

Recently, a two-step fabrication strategy, combining EBL and reactive ion etching (RIE) with a subsequent anisotropic wet etching process, to fabricate TMD metasurfaces with atomically precise edges, was developed~\cite{munkhbat2020transition,zograf2025ultrathin}. This anisotropic wet etching has provided significant improvement over conventional dry-etch-based nanofabrication. In this wet etching process, lithographically defined nanoholes evolve into energetically favored atomically sharp shapes guided by the stacking symmetry of the crystal: multilayer 3R-TMD materials preferentially form triangular features (with 60\textdegree\ internal angles) and multilayer 2H-TMD materials preferentially form hexagonal features (with 120\textdegree\ internal angles). While this phase-dependent shape selectivity ensures atomically precise and crystallographically aligned edges, it simultaneously imposes a fundamental geometric limitation: within the wet-etching time window required to form stable triangular or hexagonal features, isolated openings relax toward energetically favored local crystallographic morphologies rather than evolving into extended 180\textdegree-aligned edges. As a result, the etched features remain isolated and lack the 180\textdegree aligned, parallel edge configurations required for extended nanophotonic elements such as gratings, waveguides, and photonic cavities. Therefore, overcoming this limitation requires a fabrication strategy that preserves crystallographic edge stability while supporting the controlled propagation and merging of etched boundaries into extended edges. This phase-dependent evolution is schematically illustrated in Fig.~\ref{fig:1}a,b, which depicts the conversion of isolated lithographically defined nanoholes into atomically sharp triangular or hexagonal geometries, as dictated by the stacking of the vdW crystal.

Here, we report the fabrication of atomically precise, crystallographically aligned nanophotonic platforms in 2H-\ce{WS2} and 2H-\ce{MoS2}, such as 1D waveguides, gratings, photonic crystals, and zone plates. By lithographically defined nanoholes aligned to the crystallographic axes, followed by dry etching and anisotropic wet etching, we unlock a previously inaccessible 180\textdegree\ edge-propagation pathway in TMDs, enabling deterministic control over etch front directionality (see Fig.~\ref{fig:1}c). This directional merging mechanism produces continuous, atomically precise sidewalls and ultrahigh-aspect-ratio nanoribbons with sub-10~nm thicknesses, unattainable by conventional top-down nanofabrication. In addition to the characterization of the fabrication quality, we also experimentally verify the role of atomic sharpness by performing back-focal plane reflection spectroscopy measurements on 1D periodic, 2H-stacked \ce{WS2} gratings. These optically rich gratings simultaneously support symmetry-protected bound states in the continuum (SP-BICs) and strong light-matter coupling with excitons in \ce{WS2} multilayers.

Importantly, this lithography-guided anisotropic etching approach provides crystallographic precision, scalable pattern transfer, and universal applicability to diverse TMD geometries, enabling the realization of advanced nanophotonic and flexible nanoelectronic structures with minimal scattering losses. Together, these capabilities establish a powerful fabrication method for engineering edge-defined photonic devices and open new opportunities in integrated nonlinear, quantum, and reconfigurable photonics based on layered vdW materials.

\section{Results}\label{sec2}
\subsection{Fabrication process}\label{subsec2}

A three-step fabrication process, previously described in detail~\cite{munkhbat2020transition,agrawal2025humidity,dewambrechies2023enhanced}, is briefly presented here. Earlier studies demonstrated the fundamental mechanism by which initial circular features transform into atomically sharp and crystallographically aligned hexagonal or triangular geometries.

Here, we expand this strategy by engineering linear nanohole arrays that open access to 180\textdegree\ aligned features, instead of using isolated nanoholes. The fabrication process used in this work consists of the following sequential steps. The first step is pattern design using lithography: initial nanoholes are defined in the e-beam resist layer covering exfoliated TMD flakes. In this stage, we vary the geometrical parameters of the sample, including the position of nanohole arrays, the center-to-center spacing between nanoholes in a linear chain, spacing between neighboring nanoholes, and the pitch between adjacent linear arrays. 

Following resist development in the second step, the lithographic pattern is transferred from the resist into the TMD layers by dry reactive ion etching (RIE) process. Here, we also examine the influence of RIE etching rate on nanoholes of different diameters, as the etch rate varies with feature size. The third and final step is anisotropic wet etching, in which the patterned nanoholes in TMDs of various radii are etched anisotropically to obtain ultra smooth nanophotonic structures. The schematic of the fabrication process is shown in Supplementary Fig. \ref{fig:S1}.

Notably, atomically sharp photonic geometries with grating gaps of sub-hundreds nm are of particular importance and at the same time are challenging to fabricate. This relevance is illustrated by our Rigorous Coupled Wave Analysis (RCWA) simulations, as well as previous experimental works, which predict enhanced optical confinement and stronger light-matter interactions in such systems~\cite{choi2025unidirectional,peng20253r,zotev2025nanophotonics}. Motivated by these challenges, we first focus on the fabrication and characterization of gratings with gaps approaching sub-hundred nm. We optimized the fabrication process in terms of the nanohole radius, the center-to-center spacing between neighboring holes, and the pitch. Specifically, we varied the nanohole radius from 20 nm to 50 nm with a pitch of 250 nm and examined the resulting grating width, gap, and etch depth after anisotropic wet etching (Supplementary Fig.\ref{fig:S2}). Although smaller nanoholes with a nominal radius of 10 nm were also investigated, their reliable opening and transfer were more sensitive to resist development, proximity effects, RIE conditions, and affecting the reproducible fabrication; therefore, 20-nm-radius nanoholes were selected as the practical lower limit for reproducible fabrication (Supplementary Fig.~\ref{fig:S11}).

Based on the fabrication optimization and RCWA simulations, we identified nanoholes of 20 nm for future fabrication. We then systematically varied the center-to-center spacing between adjacent nanoholes, from 20 to 40 nm, while keeping the hole radius fixed, to evaluate lithography-induced proximity effects (Supplementary Fig.\ref{fig:S3}). At these small separations, cumulative electron scattering during EBL leads to dose enhancement and feature broadening, which is further amplified during RIE. Proximity-effect correction was intentionally not applied, as it introduces spatially varying dose distributions that result in non-uniform grating gaps across the grating. Moreover, for center-to-center spacing of 40 nm with an initial nanoholes radius of 20 nm, electron-beam proximity effects are minimized.

\begin{figure}
    \centering
    \includegraphics[width=0.99\linewidth]{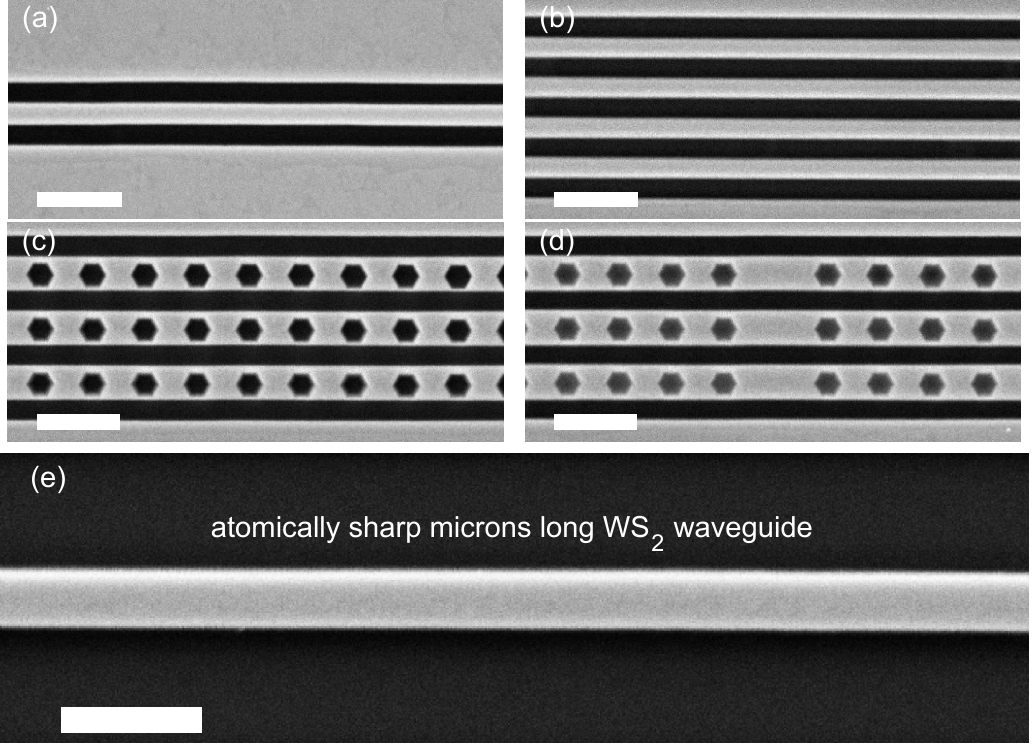}
    \caption{\textbf{Atomically sharp 180\textdegree\ aligned \ce{WS2} nanophotonic structures enabled by lithography-guided anisotropic etching.} \textbf{a} SEM image of an isolated 1D \ce{WS2} waveguide exhibiting uniform width ($\sim$107 nm) and a well-defined etched gap ($\sim$103 nm), with straight, atomically sharp sidewalls along the entire waveguide length (limited by the flake size). \textbf{b} SEM image of a periodic 1D \ce{WS2} grating composed of parallel waveguide elements with uniform widths ($\sim$107 nm) and gaps ($\sim$103 nm), demonstrating long-range dimensional uniformity. \textbf{c} Defect-free periodic \ce{WS2} photonic cavity formed by modifying the grating geometry, showing uniform grating width ($\sim$210 nm) and gap ($\sim$100 nm) with atomically sharp edges across the patterned region. \textbf{d} Defect-engineered \ce{WS2} photonic cavity with a localized perturbation into the periodic lattice; the defect is precisely defined and transferred while preserving edge sharpness and lattice uniformity. \textbf{e} SEM image of an isolated long \ce{WS2} waveguide extending over micrometer length scales, highlighting continuous, atomically sharp zigzag-enriched edges. Scale bar for all images is 500 nm.}
    \label{fig:2}
\end{figure}

Fig.~\ref{fig:2} summarizes various exemplary \ce{WS2} photonic geometries obtained in this way. Specifically, our approach allows fabricating ultra smooth waveguides, gratings, and 1D photonic crystals. The latter can be additionally engineered to have or not to have defects. Fig.~\ref{fig:2}a shows an isolated 1D \ce{WS2} waveguide with a uniform width of 107 nm and a well-defined gap of 103 nm which separates it from the rest of the unstructured \ce{WS2} flake. The nanoholes radius to fabricate these photonic geometries is 20 nm with a center-to-center spacing of 40 nm. The waveguide maintains a constant width along its entire length (limited by the flake size) with straight, well-defined edges that exhibit no observable line-edge roughness. This fabrication method is consistent and produces long, uninterrupted, atomically sharp zigzag edges governed by the underlying crystal symmetry of the initial exfoliated \ce{WS2} flake.

Building on the waveguide approach, more complex geometries can be obtained by applying the method repeatedly. Fig.~\ref{fig:2}b shows a periodic \ce{WS2} grating fabricated in an analogous way. The grating exhibits a highly uniform structure with gaps of 103 nm and consistent feature widths of 107 nm, with all edges remaining strictly parallel to each other over micrometer-length scales (defined by the crystal structure of the initial \ce{WS2} flake). Most notably, the grating edges appear straight, continuous, and free of observable line-edge roughness, indicating atomically sharp and uniform edges across the entire periodic structure. The combination of atomically sharp and uniform edges, sub-100-nm grating gaps, and long-range dimensional uniformity observed in these structures is particularly important for developing defect-free nanophotonic devices. Moreover, the ability to reproducibly fabricate both isolated and periodic geometries within the same material platform enables flexible device design, including mode-guiding, diffraction-based coupling, and resonant field enhancement. As a result, these atomically defined \ce{WS2} nanostructures represent a promising building block for next-generation integrated nanophotonic and light–matter interaction systems.

Fig.~\ref{fig:2}c and~\ref{fig:2}d demonstrate the extension of the fabrication strategy towards edge-defined photonic cavity structures of grating width 210 nm and gap 100 nm, realized through controlled modification of periodic grating geometries. Fig.~\ref{fig:2}c shows a defect-free periodic photonic cavity, where uniform grating gaps and straight, atomically sharp edges form a highly ordered cavity structure across the patterned region. The absence of missing or distorted features highlights the robustness of the anisotropic etching process even in complex, densely patterned cavity geometries. In contrast, Fig.~\ref{fig:2}d presents a defect-engineered photonic cavity, where a localized perturbation is intentionally introduced within the periodic structure. This defect is defined in the EBL step, resulting in a controlled break in periodicity while preserving atomically sharp and uniform edges throughout the nanophotonic structure. The direct comparison between Fig.~\ref{fig:2}c and~\ref{fig:2}d demonstrates that our approach enables deterministic control over both periodic and defect-defined photonic structures within the same material platform. The ability to introduce localized defects without compromising edge quality or long-range uniformity is a key structural requirement for photonic cavity designs and provides a solid foundation for future investigations of localized optical modes and enhanced light-matter interactions. Additional atomic force microscopy (AFM) analysis of the presented structures (waveguides, gratings, and photonic crystal cavities) is presented in Supplementary Fig. \ref{fig:S4}. Finally, in Fig.~\ref{fig:2}e, we present the SEM image of the isolated long \ce{WS2} grating, which is atomically sharp and zigzag edge-enriched.

To further assess the robustness of the lithography-guided anisotropic etching strategy, we systematically investigated the influence of the initial nanohole radius on the formation of edge-defined \ce{WS2} gratings. Supplementary Fig.\ref{fig:S5} shows representative SEM images of gratings fabricated using nanoholes with a radius of 500 nm, 250 nm, 100 nm, and 25 nm, while maintaining identical crystallographic alignment and etching conditions. For large nanohole radii (500 nm and 250 nm), the resulting gratings exhibit well-defined periodic gaps and straight edges, reflecting the dominance of anisotropic wet etching in shaping the final geometry. As the nanohole radius is reduced to 100 nm and further to 25 nm, the grating structures still remain continuous and crystallographically aligned, demonstrating that the anisotropic etch process reliably propagates and merges etch fronts even when initiated from deeply sub-100-nm features. Taken together, these results demonstrate that the formation of atomically sharp, crystallographically aligned \ce{WS2} gratings is largely insensitive to the initial nanohole radius across a broad range of feature sizes. This robustness indicates that the final grating geometry is governed primarily by anisotropic etch-front propagation and merging, rather than by the absolute size of the lithographically defined nanoholes.


\subsection{Optical characterization of ultrasmooth grating structures}

\begin{figure}
    \centering
    \includegraphics[width=0.99\linewidth]{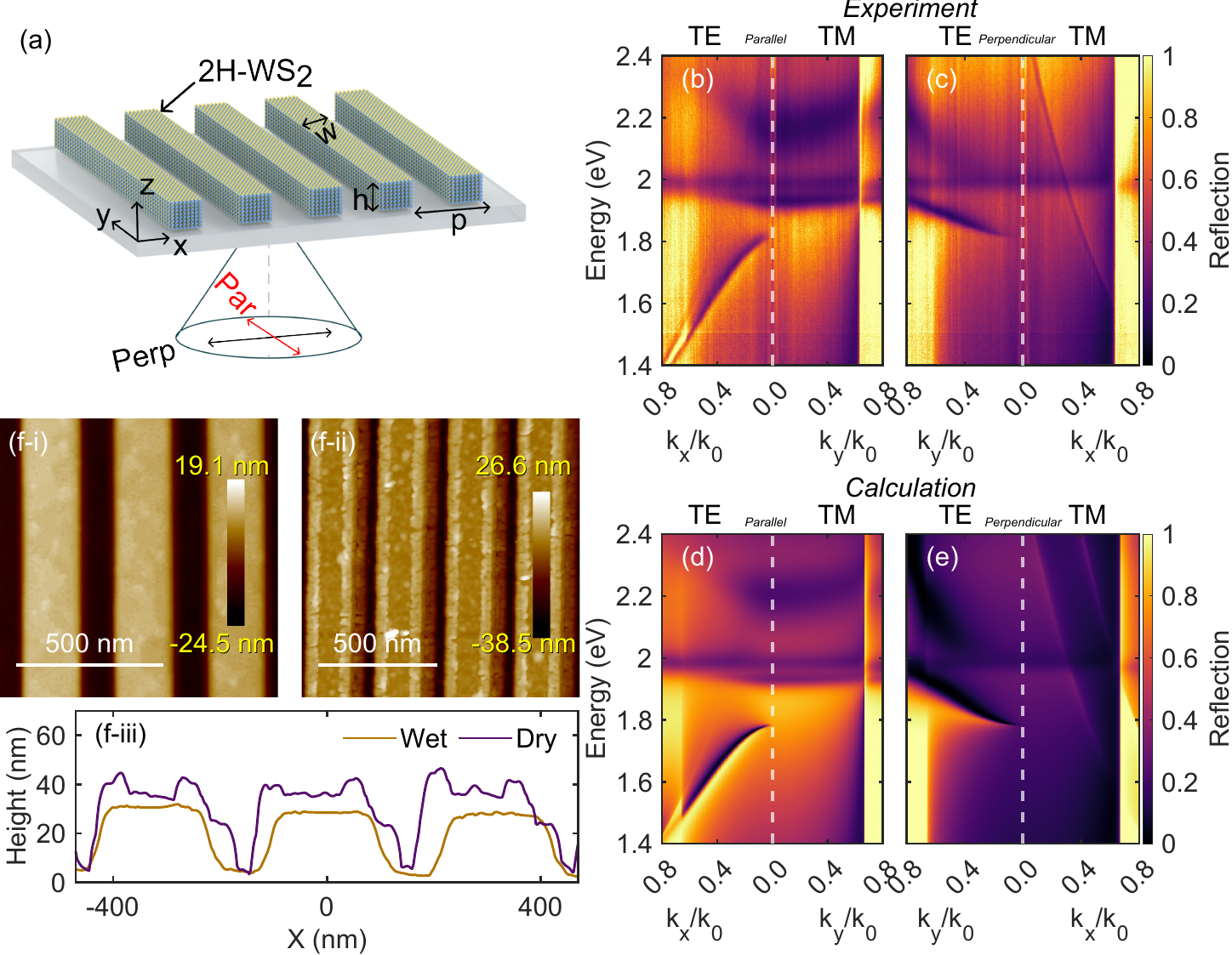}
    \caption{\textbf{Optical characterization of the periodic 1D \ce{WS2} grating.} \textbf{a} Schematic illustration of the different excitation configurations parallel and perpendicular to the grating ridges. \textbf{b} Measured dispersion spectra obtained under TE and TM excitation parallel to the grating ridges. \textbf{c} Measured dispersion spectra obtained under TE and TM excitation perpendicular to the grating ridges. \textbf{d} Calculated dispersion spectra obtained under TE and TM excitation parallel to the grating ridges, corresponding to the measurement in \textbf{b}. \textbf{e} Calculated dispersion spectra obtained under TE and TM excitation perpendicular to the grating ridges, corresponding to the measurement in \textbf{c}. \textbf{f} $2$D atomic force microscopy height maps of the (i) wet etched (sharp) and (ii) dry etched (rough) periodic grating structure. Cross-sectional measurements of the grating height corresponding to the white dashed lines are shown in (iii).}
    \label{fig:3}
\end{figure}

To assess the applicability of these lithography-guided, anisotropically-etched structures for advanced nanophotonics applications, we perform angle-resolved reflection spectroscopy on the 1D periodic \ce{WS2} grating using a Fourier plane imaging setup with excitation from the substrate side, as presented in Fig.~\ref{fig:3}. The reflection spectra are recorded in the back focal plane (BFP) of an oil-immersed $60\times$ objective (NA = $1.49$), enabling access to large in-plane wave vectors and thus to optical modes below the light line (see Methods for details). As the grating exhibits geometrical anisotropy, four distinct excitation polarization configurations can be defined. Fig.~\ref{fig:3}a defines these polarizations as either parallel or perpendicular with respect to the grating ridge direction. For both orientations, transverse electric (TE) and transverse magnetic (TM) excitation are characterized. The periodic gratings have a height of $h = 30$ nm, a width of $w = 200$ nm, and a pitch of $p = 300$ nm, as obtained from AFM measurements. 

The geometrical dimensions of the grating are designed to promote hybridization between the supported photonic modes and the intrinsic A-exciton resonance of \ce{WS2}, thereby reaching the so-called self-hybridization regime~\cite{munkhbat_self-hybridized_2019,weber_intrinsic_2023}. Fig.~\ref{fig:3}b shows the measured reflection spectra for TE (left side) and TM (right side) excitation polarized parallel to the grating ridges. Under TE excitation, the structure sustains two distinct optical modes that hybridize near the $\Gamma$ point via diffractive mode coupling. Here, the upper branch inherits all radiative loss, whereas the lower branch becomes non-radiative and evolves into a symmetry-protected bound state in the continuum (SP-BIC) near $1.8$~eV. BICs, originally introduced in the context of quantum mechanics~\cite{VonNeumann}, are special modes that reside within the radiation continuum, yet remain decoupled from it. Consequently, they can not be excited by plane waves and are invisible in far-field experiments. In the ideal case, BICs are radiationally lossless resonances characterized by diverging quality factors ($Q$-factor) and vanishing linewidths \textit{i.e.} an infinite lifetime~\cite{hsu_bound_2016}. Away from the $\Gamma$-point, the $C_2$ symmetry is broken, enabling radiative coupling and transforming the BIC into a quasi-BIC with a finite $Q$-factor, as clearly observed in Fig.~\ref{fig:3}b. Strong interaction between the photonic mode and the \ce{WS2} A-exciton ($\sim 2$~eV) is observed in the upper optical branch, splitting into the lower and upper polariton (UP) bands. These new eigenmodes of the hybridized system are quasi-particles that are both light- and matter-like, inheriting properties from both components. Under TM excitation parallel to the ridges (right side), the LP band flattens out, resulting in a nearly dispersionless polaritonic branch. Such flat, self-hybridized dispersion is a promising platform for optoelectronic applications, enabling broadband absorption over a large range of in-plane wave vectors, sub-bandgap absorption, and enhanced exciton diffusion~\cite{munkhbat_self-hybridized_2019,jariwala_near-unity_2016,wong_high_2017}. The highly reflective region beyond the critical angle ($k_y/k_0 \sim 0.66$) corresponds to total internal reflection from the glass/air interface. The dispersion spectra for the two remaining polarization configurations (perpendicular to the grating ridges) are shown in Fig.~\ref{fig:3}c. Under TE illumination, a SP-BIC appears at the $\Gamma$-point and evolves into a quasi-BIC at oblique incidence. As the quasi-BIC disperses toward the excitonic resonance, it shows an avoided crossing leading to the formation of LP and UP bands. In contrast, under TM excitation perpendicular to the ridges, only the exciton absorption and the onset of a diffraction order are visible in the dispersion, attributable to the limited thickness of the \ce{WS2} flake ($\sim 30$ nm). RCWA calculations corresponding to the measurements in Fig.~\ref{fig:3}b and \ref{fig:3}c are shown in Fig.~\ref{fig:3}d and \ref{fig:3}e, respectively. The calculations employ a dispersive refractive index of \ce{WS2} obtained from ellipsometry~\cite{munkhbat_optical_2022} and identical geometrical parameters obtained from AFM measurements. The measurements and calculations show excellent agreement, indicative of sample fabrication with well-defined, sharp features. Lastly, Fig.~\ref{fig:3}f presents 2D AFM height maps of (i) the anisotropically wet-etched grating and (ii) a benchmark grating with dimensions defined solely by the dry etching step. For the wet-etched sample, a uniform grating modulation can be observed across the entire scanned area. In contrast, the dry-etched grating shows a more inhomogeneous thickness with rougher sidewalls. This observation is confirmed in Fig.~\ref{fig:3}f(iii), showing cross-sections of the height of the gratings along the $x$-direction corresponding to the dashed lines in the AFM maps. We conclude that the wet etching step improves the geometry of the gratings in several ways. First, the grating gap channels, with widths of approximately $100$ nm, are opened, thereby significantly reducing the tapering of the grating sidewalls. Second, wet etching removes the excess material accumulated at the top of the grating ridges. In the benchmark structure, repeating crown-like features appear near the ridge edges, most likely originating from material re-deposition during the dry etching process. As these features are typically unstable, an additional wet-etching step removes them, cleaning the top of these surfaces.

In addition to the significant improvements achieved in fabrication, we compare the optical performance by characterizing the response of the dry-etched gratings in the BFP. The resulting dispersion spectra are presented in Fig.~\ref{fig:S5_5}a,b for parallel and perpendicular polarization excitation, respectively. Apart from a slight blue shift, attributable to a combination of grating wall tapering, fabrication roughness, and slight deviations in the grating dimensions, only minor differences are observed between the sharp and benchmark gratings under identical excitation configurations. Due to the limited angular and spectral resolution of the setup, both grating structures exhibit comparable $Q$-factors. Therefore, we analyze the effect of surface roughness by considering high-$Q$ modes in FDTD simulation. Fig.~\ref{fig:S5_5}c provides a direct comparison of the $Q$-factors for a high-$Q$ quasi-BIC mode, showing a significant decrease in $Q$-factor when artificial roughness is added on the top facets of the gratings (corresponding to the dry-etched structure) as compared to a perfectly sharp modulation. Additionally, a reduced optical contrast and unaccounted blueshift are observed for the roughened sample, marking the importance of precise fabrication in preserving the designed optical performance. Finally, our method extends beyond the fabrication of gratings. By only partially etching the \ce{WS2} flakes, thereby leaving an unetched waveguide slab underneath a periodic grating modulation, structures sustaining guided mode resonances (GMRs) are realized. Fig.~\ref{fig:S5_6} summarizes the optical response of these structures, presenting both measured and simulated angle-resolved reflection spectra alongside the calculated field distributions, demonstrating the presence of BICs and GMRs.

\subsection{Advanced two-dimensional structures}

Anisotropic wet etching in layered 2H-phase TMDs is governed by crystal stacking symmetry and therefore produces discrete, crystallographically aligned edge orientations rather than disordered structures. In isolated geometries such as circles, this anisotropy stabilizes zigzag edges that intersect at 120\textdegree\, leading to hexagonal shapes. By lithographically enforcing the spatial arrangement of etch initiation sites, the present approach also enables coherent merging of opposing zigzag etch fronts. As a result, extended straight edges can be formed by combining intrinsic 120\textdegree\ junctions with lithographically enabled 180\textdegree\ segments, producing complex photonic structures composed exclusively of integer combinations of 120\textdegree\ and 180\textdegree\ angles.

\begin{figure}
    \centering
    \includegraphics[width=0.99\linewidth]{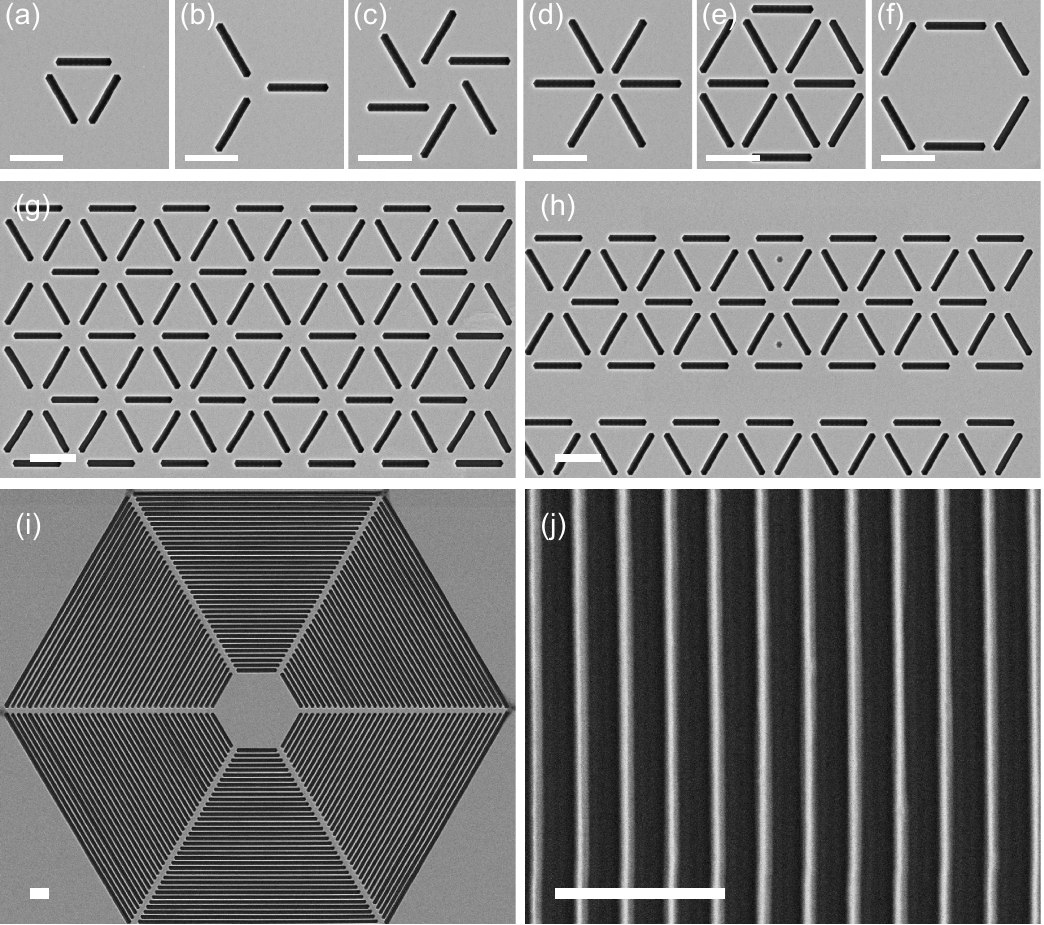}
    \caption{\textbf{Angle-selective edge engineering in multilayer 2H-\ce{WS2} enabled by anisotropic etching.}  \textbf{a-f} SEM images of isolated and coupled edge-defined photonic structures composed of linear segments connected exclusively through crystallographically allowed 120\textdegree\ and lithographically enabled 180\textdegree\ edge junctions. Despite increasing geometric complexity, all edges remain straight, uniform, and atomically sharp, indicating that local edge formation is governed by crystallographic anisotropy while global geometry is programmed by controlled etch-front. \textbf{g-h} Periodic edge-defined photonic lattices formed by tiling triangular units into extended two-dimensional networks, resembling photonic-crystal-like structures defined by discrete 120\textdegree\ junctions and continuous 180{\textdegree}-aligned edges. Edge quality and angular fidelity are preserved across multiple unit cells without observable accumulation of line-edge roughness. \textbf{i} Radially patterned hexagonal zone-plate geometry constructed from concentric, crystallographically aligned grating segments defined entirely in multilayer radius \ce{WS2}. \textbf{j} Magnified view of the zone plate, showing sub-100-nm grating gaps, long-range straightness, and atomically sharp edges across the aperiodic structure. Scale bar for all images is 1~$\mu$m. }
    \label{fig:4}
\end{figure}

Fig.~\ref{fig:4} demonstrates how the angle-selective edge-formation mechanism introduced above can be applied to fabricate complex photonic structures composed exclusively of crystallographically allowed 120\textdegree\ and 180\textdegree\ edge connections. By combining intrinsic zigzag junctions at 120\textdegree\ with lithographically enabled 180\textdegree\ aligned segments, the present fabrication strategy enables deterministic construction of 2D geometries that extend beyond isolated hexagonal shapes.

Fig.~\ref{fig:4}a-f present a series of isolated and coupled edge-defined photonic structures, in which linear edge segments are connected through combinations of 120\textdegree\ and 180{\textdegree}. These photonic structures include triangular and polygonal arrangements, as well as more complex coupled configurations where multiple crystallographic directions coexist within a single architecture. Despite increasing geometric complexity, all edge segments remain straight, uniform, and atomically sharp, indicating that edge formation is governed locally by crystallographic symmetry, while the global geometry is programmed through lithographic control of etch-front initiation and merging.

Fig.~\ref{fig:4}g and~\ref{fig:4}h extend this angle-selective design strategy to periodic edge-defined photonic lattices, formed by tiling triangular photonic structures into extended 2D networks. The resulting structures resemble photonic-crystal-like lattices in which periodicity and angular connectivity are dictated by discrete 120\textdegree\ junctions and continuous 180\textdegree\ edges. Importantly, edge quality and angular fidelity are preserved across many unit cells, with no observable accumulation of line-edge roughness or angular distortion, demonstrating that coherent etch-front propagation and merging are maintained over extended areas.

Fig.~\ref{fig:4}i and~\ref{fig:4}j demonstrate that the angle-selective edge-engineering strategy can be extended beyond periodic lattices to realize functionally graded diffractive photonic structures. Fig.~\ref{fig:4}i shows a radially patterned hexagonal zone-plate geometry, constructed from concentric segments defined by crystallographically aligned, edge-defined gratings. In contrast to conventional zone plates fabricated in low-index dielectrics or plasmonic materials, this structure is realized entirely in multilayer \ce{WS2} and defined by atomically sharp edges constrained to crystallographically allowed orientations through anisotropic etching. To the best of our knowledge, such an edge-defined zone plate implemented in a TMD platform has not been previously reported. Fig.~\ref{fig:4}j presents a magnified view of the zone-plate structure, revealing that the individual grating segments preserve sub-100-nm gaps, long-range straightness, and atomically sharp edges across the entire radially varying pattern. The uniform edge quality observed despite continuous changes in local grating orientation confirms that the angle-selective edge-formation mechanism, based on intrinsic 120\textdegree\ junctions and lithographically enabled 180\textdegree\ segments, remains robust in aperiodic and non-uniform geometries. We also show patterned zone-plate geometry in 2H-\ce{MoS2} in Supplementary Fig. \ref{fig:S8}. This result establishes that the present fabrication strategy supports not only periodic photonic lattices but also aperiodic, phase-engineered diffractive elements, significantly broadening the scope of edge-defined nanophotonic design.

Finally, we further explore other possible angles, such as 240\textdegree\ and 300\textdegree\ in 2H-\ce{WS2} and 3R-\ce{MoS2}, respectively. Supplementary Fig.\ref{fig:S9} examines structures designed to enforce 240\textdegree\ and 300\textdegree\ angular configurations, corresponding to positive hexagonal and triangle geometries. They are not stable under anisotropic wet etching and do not preserve their intended shape during continuous wet etching. The edges of the positive hexagons and positive triangles are not sharp and are unstable. Moreover, structures defined by a larger initial hole radius require longer wet-etch times to reach crystallographic equilibrium.

\subsection{State-of-the-art aspect ratio nanoribbons}

\begin{figure}
    \centering
    \includegraphics[width=0.99\linewidth]{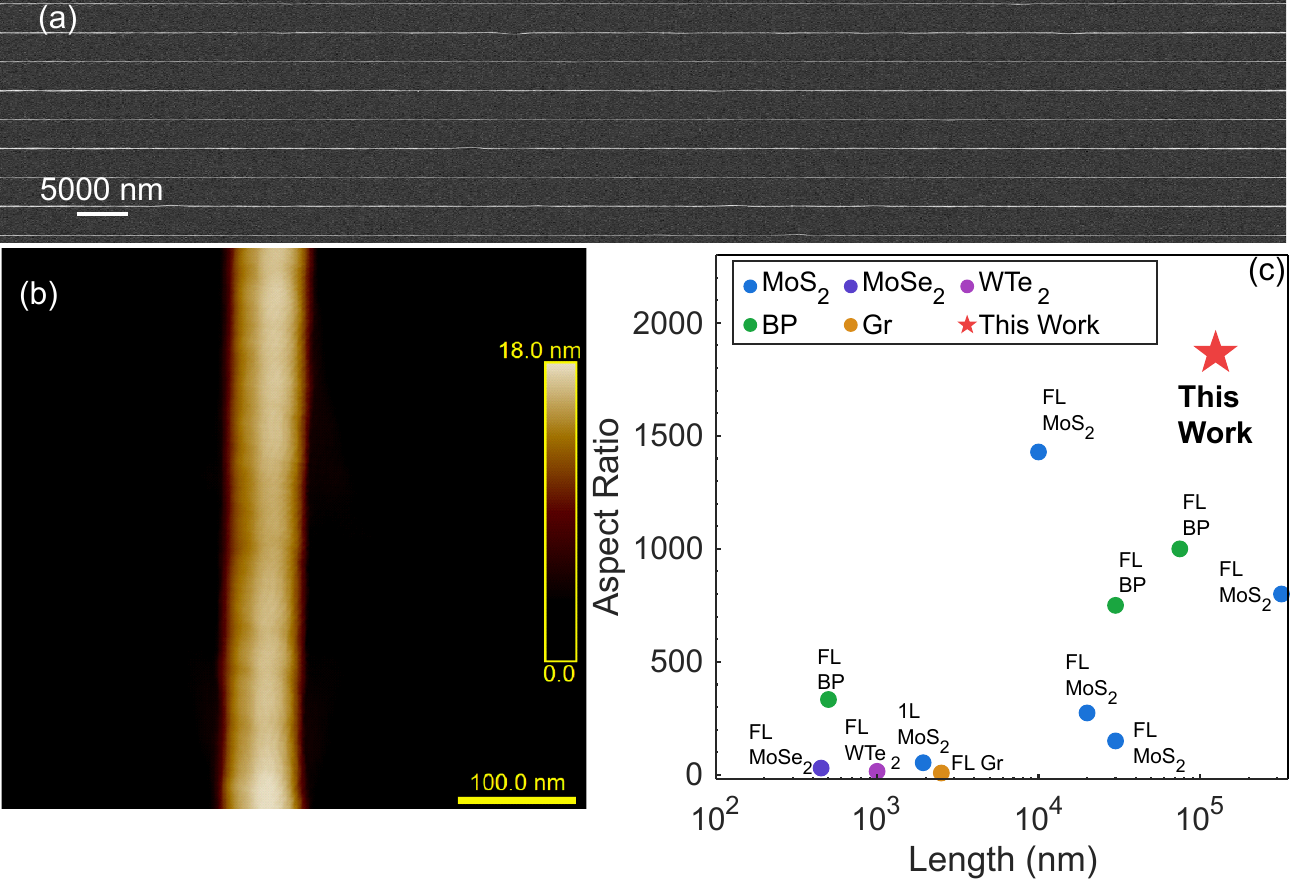}
    \caption{\textbf{Ultralong, ultrathin \ce{WS2} nanoribbons enabled by sustained 180\textdegree\ edge propagation.} \textbf{a} Large-area SEM image showing multiple parallel \ce{WS2} nanoribbons extending continuously over lengths of $\sim$ 50 $\mu$m, demonstrating record aspect ratio and uniform width enabled by lithography-guided anisotropic etching. Scale bar of the SEM image is 5 $\mu$m. \textbf{b} Atomic force microscopy (AFM) height map of an individual nanoribbon, revealing an ultrathin thickness of $\sim$ 18 nm and a laterally uniform width of $\sim$ 67 nm across the ribbon. \textbf{c} Comparison of nanoribbon aspect ratio (length/width) as a function of length for TMD nanoribbons fabricated using various synthesis methods. The references can be found in Supplementary Information and in Table S1.}
    \label{fig:5}
\end{figure}

Fig.~\ref{fig:5} presents an extended demonstration of the present fabrication approach, showing that the same lithography-guided anisotropic etching strategy can be used to produce ultrathin, ultranarrow, and ultralong \ce{WS2} nanoribbons with record aspect ratios. Fig.~\ref{fig:5}a shows a large-area SEM image of multiple nanoribbons that continuously extend over lengths of approximately 50 $\mu$m. The nanoribbons remain straight and parallel throughout the imaged region, with no visible interruptions or width modulation, indicating stable pattern transfer and uniform edge evolution during extended etching. Currently, the nanoribbon length achievable is primarily limited by the lateral size of the exfoliated \ce{WS2} flakes rather than by the fabrication process itself. This suggests that further increases in the length of the nanoribbon may be achievable by employing larger areas \ce{WS2} flakes or wafer-scale material platforms.

Fig.~\ref{fig:5}b shows an AFM height profile of an individual nanoribbon. The measured height is approximately 18~nm, with a laterally uniform thickness profile across the ribbon width of $\sim 67$~nm. This confirms that the nanoribbons are ultrathin and laterally well confined, while maintaining structural continuity over tens of micrometers. The combination of nanoscale thickness and width with micrometer-scale length results in a highly elongated geometry that is difficult to achieve using conventional growth- or exfoliation-based approaches. Fig.~\ref{fig:5}c compares the aspect ratio (length/width) as a function of length for TMD nanoribbons fabricated via solution synthesis, chemical vapor deposition (CVD), molecular beam epitaxy (MBE), pulsed laser deposition (PLD), chemical vapor transport (CVT), and exfoliation-based fabrication. The data point corresponding to this work lies at both a larger length and a higher aspect ratio than the majority of previously reported structures, highlighting that the present method enables access to a regime of ultrathin, ultranarrow, ultralong nanoribbons, a regime that has been sparsely explored. Beyond their exceptional aspect ratios, these nanoribbons are distinguished by their atomically sharp and crystallographically aligned zigzag edges, enabled by sustained 180° etch-front propagation in multilayer 2H-\ce{WS2}. To the best of our knowledge, such ultralong TMD nanoribbons with simultaneously high aspect ratio, long-range edge uniformity, and crystallographically defined atomic edge alignment have not been previously realized.

Taken together, these results show that lithography-guided anisotropic etching can reproducibly generate \ce{WS2} nanoribbons combining sub-20-nm thickness, sub-100-nm width, and tens-of-micrometers length. Such geometries are of interest for nanophotonic and optoelectronic applications where long-range structural uniformity and reduced edge disorder are important. Supplementary Fig.\ref{fig:S10} provides additional SEM and AFM characterization of extended \ce{WS2} nanoribbons with lengths up to 125~$\mu$m. AFM measurements show ribbon thicknesses in the range of 12 -- 22~nm and lateral widths of 40 -- 70~nm, with uniform profiles along the ribbon length. These results further support the formation of ultrathin, ultranarrow, and ultralong nanoribbons, with the maximum length primarily limited by the lateral size of the exfoliated \ce{WS2} flakes.

\section{Discussion}\label{sec10}

The fabrication of atomically sharp gratings, waveguides, photonic cavities, and ultralong nanoribbons is achieved by systematically controlling the nanohole radius, nanohole placement, center-to-center nanohole spacing, and nanohole array pitch during lithographic patterning, followed by a combination of dry etching and anisotropic wet etching. This fabrication strategy enables atomically sharp gratings and waveguides Fig.~\ref{fig:2}, angle-resolved photonic geometries Fig.~\ref{fig:4}, and ultranarrow, high–aspect-ratio nanoribbons Fig.~\ref{fig:5} in multilayer 2H-TMDs. This work establishes lithography-guided anisotropic wet etching as a general strategy to overcome the geometric constraints that have previously limited atomically precise nanostructures of TMDs to isolated shapes. By enforcing controlled interaction between neighboring etched regions, the fabrication approach enables coherent propagation and merging of crystallographically aligned stabilized edges, introducing a previously inaccessible 180\textdegree\ edge alignment. This additional degree of freedom transforms isolated atomic metasurfaces, such as hexagons and triangles, into extended and interconnected nanophotonic structures. 

Wet etchants such as peroxide-based solutions or dry etching with oxygen preferentially react with the TMD's basal plane in all directions, but etch fastest along armchair edges, therefore only zigzag-terminated edges remain once the process is finalized. Each zigzag edge may terminate with either metal atoms (M-ZZ) or chalcogen atoms (X-ZZ). \textit{Ab initio} and experimental studies reveal that under oxidizing or basic conditions, M-ZZ edges possess lower formation energies and higher chemical stability than X-ZZ edges~\cite{dewambrechies2023enhanced,li2024anisotropic,munkhbat2020transition,lauritsen2007size,yang2010anisotropic}. Consequently, X-ZZ edges etch faster, and initially circular or irregular holes gradually evolve into triangular pits bounded exclusively by the more stable termination. These triangular holes correspond to a single zigzag orientation with 60\textdegree\ internal angles, reflecting one dominant termination (often Mo-ZZ in \ce{MoS2}). In monolayer 2H–\ce{MoS2} or \ce{WS2}, only one MS$_2$ layer is present, and thus no alternating edge termination exists. Etching, therefore, yields equilateral triangular holes, since all edges share the same in-plane orientation and termination~\cite{munkhbat2020transition}. The same triangular geometry is observed in 3R-phase \ce{MoS2}, where all layers are aligned without 180\textdegree\ rotation typical for AB stacking of 2H-phases, maintaining a single zigzag type through the stack, as shown in  Fig.~\ref{fig:1}a. In both cases, monolayer 2H and multilayer 3R, the absence of inversion symmetry and alternating terminations confines etching to one energetically favored direction, producing triangular nanoholes~\cite{zograf2025ultrathin}. By contrast, multilayer 2H-phase (AB-stacked) TMDs exhibit inversion symmetry, with each adjacent layer rotated by 180\textdegree. This arrangement alternates M-ZZ and X-ZZ terminations vertically, allowing both edge types to coexist and stabilize during etching. As a result, the etched structures display hexagonal holes with 120\textdegree\ internal angles, bounded by alternating zigzag edges, as shown in 
Fig.~\ref{fig:1}b~\cite{munkhbat2020transition}. However, if stacking faults or 3R-like domains occur within a 2H crystal, the alternation is disrupted, and the etching reverts to a single termination, locally forming triangular holes~\cite{dewambrechies2023enhanced,polyakov2024top}. The shape evolution of an isolated drilled hole is primarily governed by the intrinsic stacking symmetry of the crystal, producing triangles (zigzag orientation with 60\textdegree\ internal angles) holes in 3R-phase and hexagonal (zigzag orientation with 120\textdegree\ internal angles) holes in 2H-phase TMDs, with size limitation defined by the radius of the drilled hole.

To overcome this limitation, we extend the anisotropic etching approach by lithographically enforcing the spatial arrangement of holes along the crystallographic zigzag axes. The key distinction introduced in this work is that the anisotropic etching process is deliberately driven into a regime where neighboring etch fronts interact before individual features reach their equilibrium termination by carefully defining the position, spacing, and orientation of the EBL-patterned nanoholes, effectively creating 180\textdegree{}-aligned features.  In conventional anisotropic etching of isolated nanoholes, each etch front evolves independently and relaxes toward a local energetic minimum, at which point lateral propagation ceases, and vertices act as termination points. By contrast, when lithographically patterned nanoholes are placed with controlled spacing and crystallographic alignment, the advancing etched boundaries intersect while edge propagation is still active. In this regime, etch fronts are no longer governed solely by local shape relaxation but by their mutual interaction, which suppresses vertex formation and prevents premature termination. As a result, instead of closing into isolated polygons, the interacting fronts merge and continue propagating as a single, extended boundary. This transition from isolated evolution to collective, interaction-driven etching fundamentally alters the outcome of anisotropic wet etching, enabling sustained edge propagation beyond the size limits imposed by isolated nanoholes.

However, this 180\textdegree\ propagation is only achievable in 2H-stacked TMDs, where two parallel zigzag facets on opposite sides of the hexagon allow neighboring etch fronts to align and coalesce seamlessly. In contrast, 3R-stacked crystals, possessing only a single zigzag orientation, form triangular pits with one flat base and two inclined sides that meet at 60\textdegree, preventing coherent merging and continuous edge formation. As a result, only 2H-TMDs can evolve into extended nanoribbons, gratings, and photonic waveguides. This directional merging mechanism establishes a new geometric degree of freedom in edge-defined photonic metasurfaces, as illustrated conceptually in Fig.~\ref{fig:1}c. 
Geometries with internal angles of 240\textdegree\ or 300\textdegree\ require edge segments that depart from the crystallographically preferred zigzag directions. Under anisotropic wet etching, such edge orientations are energetically unstable and cannot be sustained (Supplementary Information Fig. \ref{fig:S9}). As a result, these geometries do not represent local minima in the edge-energy landscape and rapidly collapse during etching, reverting to stable 120\textdegree\ junctions or terminating altogether. The absence of stable edge terminations supporting positive curvature therefore restricts the design space to dynamically stable 120\textdegree\ (equilibrium) and 180\textdegree\ (interaction-driven) angles.

The restriction to a small set of crystallographically allowed angles is an advantage for nanophotonic design. In nanoscale photonic devices, optical losses are often dominated by scattering at rough edges, curved boundaries, and disordered junctions. Here, all structures are formed from atomically sharp zigzag edges joined only at stable angles, which naturally suppress these loss channels. Straight 180\textdegree\ edges provide low-loss waveguides and gratings, while 120\textdegree\ junctions enable controlled bending and lattice formation without introducing sharp corners or uncontrolled curvature. By eliminating arbitrary angles, this approach enforces geometric uniformity and reproducibility, allowing complex photonic structures to be built from a small number of stable building blocks, introduced in this work. The discrete, crystal-defined angle set combination of 180\textdegree\ and 120\textdegree\ enables complex periodic photonic structures, including diffractive zone plates, to be realized using only intrinsically stable edge segments. As a result, geometric complexity can be increased without compromising atomic-scale edge precision.

Ultralong, ultrathin, and ultranarrow nanoribbons represent the most extreme outcome of sustained 180\textdegree\ edge propagation enabled by lithography-guided anisotropic etching. In this regime, paired zigzag edges merge and extend together, forming ribbons with uniform width and atomically sharp sidewalls over tens to hundreds of micrometers. Unlike conventional nanoribbon fabrication, where rough edges or defects limit length, ribbon growth here is governed by edge continuity rather than lithographic resolution. The absence of edge degradation indicates that the merged zigzag boundaries remain energetically stable during prolonged etching. At present, ribbon length is mainly limited by flake size, suggesting further scaling is possible with larger-area single-crystalline materials.

Despite its versatility, the fabrication strategy has several practical limitations. The lateral size of the structures is currently limited by the dimensions of exfoliated multilayer \ce{WS2} flakes, which constrains the maximum length of waveguides, gratings, and nanoribbons. The approach also relies on the stacking symmetry of multilayer 2H-phase TMDs and is therefore not applicable to monolayer or 3R-stacked materials, where bidirectional edge stability is absent. In addition, anisotropic wet etching requires precise crystallographic alignment and careful control of etching conditions to ensure reproducible results over large areas. Overcoming these constraints will be important for extending this strategy to wafer-scale platforms and broader material systems.

The practical lower limit of the nanoholes seed geometry is governed by the ability to fully open small features during EBL resist development and to reliably transfer them into the TMD layer during RIE. We therefore evaluated 10 nm radius nanoholes with center-to-center spacings of ($r$), ($3r/2$), and ($2r$) (Supplementary Fig.\ref{fig:S11}). Although closely spaced 10-nm-radius arrays with ($r$) or ($3r/2$) spacing can initiate continuous grating formation, their effective opening is strongly influenced by collective resist development, proximity-induced broadening, and dry-etch duration. Longer RIE can improve the transfer of such dense patterns, but may also broaden the effective seed openings and modify the final grating gap after anisotropic wet etching. For this reason, 20 nm radius nanoholes with ($2r$) center-to-center spacing were selected for the main fabrication workflow, as they provide more reproducible hole opening and pattern transfer, while comparable grating gaps can still be achieved by tuning the pitch between neighboring linear nanoholes arrays. Further discussion of the practical nanoholes size and spacing limits is provided in Supplementary Note 11.

\section{Conclusion}\label{sec11}

In conclusion, we have established a crystallography-guided fabrication framework that extends anisotropic wet etching of TMDs beyond isolated shape formation. By systematically controlling lithographic parameters, including nanohole size, placement, inter-hole spacing, array pitch and combining dry etching with anisotropic wet etching, we enable controlled interaction and merging of etched boundaries. This approach introduces a previously inaccessible 180\textdegree\ edge alignment, complementing intrinsic 60\textdegree\ (in 3R-) and 120\textdegree\ (in 2H-) equilibrium geometries and enabling the deterministic formation of extended, atomically sharp nanophotonic structures in multilayer 2H-TMDs. Using this strategy, we demonstrate a broad range of edge-defined structures, including atomically sharp gratings, waveguides, defect-engineered photonic cavities, angle-resolved photonic geometries, diffractive zone plates, and ultrathin nanoribbons with exceptionally high aspect ratios. The results show that edge continuity and connectivity, rather than lithographic resolution alone, govern the achievable geometry once equilibrium termination is overcome. Importantly, the observed phase selectivity reveals that only multilayer 2H-stacked materials support sustained bidirectional edge propagation and merging, while multilayer 3R-stacked crystals remain confined to isolated equilibrium shapes. Beyond providing atomic-scale structural precision, the presented framework defines a crystallographically quantized design space, where only energetically and dynamically stable angles are preserved during etching. This constraint ensures high reproducibility and uniformity across device-relevant length scales, offering a robust structural foundation for low-scattering nanophotonic elements based on two-dimensional materials. While the present study focuses on fabrication and structural characterization, the demonstrated control over edge geometry and continuity opens clear pathways toward future investigations of integrated photonics, diffractive optics, and edge-mediated light-matter interactions in layered TMD systems.

\section{Methods}\label{sec12}
\subsection{Sample fabrication}\label{sec12-1}

Exfoliation: TMDs flakes were mechanically exfoliated from bulk crystals (HQ-graphene) onto polydimethylsiloxane (PDMS) stamps using the scotch-tape method. For standard fabrication, selected flakes were transferred onto Si substrates capped with a 285-nm-thick SiO$_2$ layer. For optical characterization, multilayer flakes with the desired thickness were transferred onto 1-inch $\times$ 1-inch glass substrates (0.17 mm thickness) using PDMS-assisted dry transfer. Electron-beam lithography (EBL) was performed using ARP 6200.13 positive resist. To mitigate charging effects during EBL on insulating glass substrates, a 25-nm-thick chromium layer was deposited on top of the resist using a Lesker Nano 36 thermal evaporator prior to exposure. Lithographic patterning was carried out with a Raith EBPG 5200 system operated at an accelerating voltage of 100~kV and a beam current of 10~nA. Following exposure, the chromium layer was removed using a Ni–Cr etchant solution (10 -- 20 \%  ammonium cerium(iv) nitrate, 5 -- 40\% nitric acid; Sunchem AB), and the resist was developed in n-amyl acetate (Sigma-Aldrich). Pattern transfer into the TMD flakes was achieved by reactive ion etching using an Oxford Plasmalab 100 (RIE/ICP) system with a CHF$_3$/Ar gas mixture. Residual resist was subsequently removed using a commercial remover, followed by thorough rinsing in deionized water and gentle nitrogen drying.

Finally, the lithographically defined nanohole arrays were transformed into atomically sharp gratings and related nanophotonic structures through anisotropic wet etching in an aqueous solution of hydrogen peroxide and ammonia~\cite{munkhbat2020transition}.   

\subsection{Atomic force microscopy} \label{sec12-2}

Atomic force microscopy (AFM) measurements were performed using a Bruker ICON system equipped with a Nanoscope~5 controller. Imaging was carried out in tapping mode using RTESP-300 probes (Bruker).

\subsection{Scanning electron microscopy} \label{sec12-3}

SEM imaging of the nanophotonics surfaces was performed at Chalmers Materials Analysis Laboratory using an Ultra~55 microscope (CarlZeiss). An acceleration voltage of 3~kV was used for imaging both samples.

\subsection{Optical reflection spectroscopy measurements} \label{sec12-4}

The dispersion spectra presented in this work are measured via a Fourier plane microscopy and spectroscopy setup. The details of steup can be found from previous reports. \cite{zograf2025ultrathin} The broadband light is generated using a laser-driven light source (LDLS) and subsequently collimated by a lens. A linear polarizer defines the polarization of the light and is employed to characterize the response of the sample incident under both transverse electric (TE) and transverse magnetic (TM) polarized excitation. The light is guided to an inverted optical microscope (Nikon Eclipse TE2000-e) and focused onto the sample by an oil-immersed objective (Nikon CFI Apo TIRF 60$\times$ Oil, MRD01691) with a numerical aperture of $\mathrm{NA} = 1.49$. This high-NA objective allows for access to large in-plane wave vectors probing optical modes below the light line (\textit{i.e.} bound modes) that would have otherwise been invisible due to their evanescent nature in air. The reflected light is either imaged onto a Nikon D300s color camera to obtain real-color images of the back focal plane (BFP), or coupled into a fiber-coupled spectrometer to measure angle-resolved reflection spectra. A phase telescope, also known as a Bertrand lens, extends the BFP of the objective to the position of a fiber bundle, consisting of 19 fibers mounted on a 3D translational stage. Since the Fourier plane encodes angular information as $r \sim \sin{\theta}$, where $r$ is the radial distance from the center and $\theta$ is the angle of incidence, the full polarization-dependent dispersion can be mapped by translating the fiber bundle across the BFP image. The spectrometer is connected to a CCD camera (Andor Newton 920) that collects the frequency-resolved light. For samples exhibiting geometrical anisotropy, such as gratings, four polarizations need to be characterized. This is done by (i) rotating the polarizer by 90\textdegree\ to probe TE and TM polarization and (ii) rotating the sample by 90\textdegree\ to change the geometrical orientation.

\textbf{Data availability} 
The data that support the findings of this study are available within the Article and its Supplementary Information. Additional data are available from the corresponding authors upon reasonable request.

\textbf{Code availability}

\backmatter

\bmhead{Supplementary information}
The online version contains supplementary material.

\bmhead{Acknowledgements}

This work was performed in part at Myfab Chalmers and Chalmers Materials Analysis Laboratory (CMAL). A.V.A, B.K, and T.O.S. acknowledge funding from the SIO Grafen VINNOVA (Ref. 2023-04142), Olle Engkvist Foundation (Grant No. 211-0063), 2D-TECH VINNOVA competence center (Ref. 2024-03852), Chalmers Area of Advance Nano, and Knut and Alice Wallenberg Foundation (KAW, grant No. 2019.0140). T.J.A. thanks the Polish National Science Center for support via the project 2019/34/E/ST3/00359.

\section*{Competing interests}
The authors declare no competing interests.

\textbf{Correspondence} and requests for materials should be addressed to Abhay V. Agrawal or Timur O. Shegai.

\bigskip

\begin{appendices}




\end{appendices}


\bibliography{sn-bibliography}

\end{document}